\documentclass[a4paper,conference]{IEEEtran}
\IEEEoverridecommandlockouts
\usepackage{cite}
\usepackage{amsmath,amssymb,amsfonts}
\usepackage{algorithmic}
\usepackage{graphicx}
\usepackage{comment}
\usepackage{textcomp}
\usepackage{xcolor}
\usepackage{tikz}
\usetikzlibrary{positioning}
\def\BibTeX{{\rm B\kern-.05em{\sc i\kern-.025em b}\kern-.08em
    T\kern-.1667em\lower.7ex\hbox{E}\kern-.125emX}}
\begin{document}

\title{Size Matters: The Impact of Avatar Size on User Experience in Healthcare Applications\\

}
\author{
\IEEEauthorblockN{Navid Ashrafi$^{1,2}$, Francesco Vona$^{1}$, Sina Hinzmann$^{1}$, Juliane Henning$^{1}$, Maurizio Vergari$^{2}$, \\ Maximilian Warsinke$^{2}$, Catarina Pinto Moreira$^{3}$, Jan-Niklas Voigt-Antons$^{1}$}
\IEEEauthorblockA{$^1$Immersive Reality Lab, Hamm-Lippstadt University of Applied Sciences, Hamm, Germany\\
$^2$Quality and Usability Lab, Technische Universität Berlin, Berlin,  Germany\\
$^3$Data Science Institute, University of Technology Sydney, Australia}
}

\maketitle

\begin{abstract}

The usage of virtual avatars in healthcare applications has become widely popular; however, certain critical aspects, such as social distancing and avatar size, remain insufficiently explored. This research investigates user experience and preferences when interacting with a healthcare application utilizing virtual avatars displayed in different sizes. For our study, we had 23 participants interacting with five different avatars (a human-size avatar followed by four, respectively, smaller avatars in a randomized order) varying in size, projected on a wall in front of them. The avatars were fully integrated with an artificial intelligence chatbot to make them conversational. Users were asked to rate the usability of the system after interacting with each avatar and complete a survey regarding trust and an additional questionnaire on social presence. The results of this study show that avatar size significantly influences the perceived attractiveness and perspicuity with the medium-sized avatars having received the highest ratings. Social presence correlated strongly with stimulation and attractiveness, suggesting that an avatar’s visual appeal and interactivity influenced user engagement more than its physical size. Additionally, we observed a tendency for gender-specific differences on some of the UEQ+ scales, with male participants tending to prefer human-sized representations, while female participants slightly favored smaller avatars. These findings highlight the importance of avatar design and representation in optimizing user experience and trust in virtual healthcare environments. 

\end{abstract}
\newcommand\copyrighttext{%
  \footnotesize \textcopyright 2025 IEEE. Personal use of this material is permitted. Permission from IEEE must be obtained for all other uses, in any current or future media, including reprinting/republishing this material for advertising or promotional purposes, creating new collective works, for resale or redistribution to servers or lists, or reuse of any copyrighted component of this work in other works. Navid Ashrafi, Francesco Vona, Sina Hinzmann, Juliane Henning, Maurizio Vergari, Maximilian Warsinke, Catarina Pinto Moreira, and Jan-Niklas Voigt-Antons. 2025. Size Matters: The Impact of Avatar Size on User
Experience in Healthcare Applications. In 2025 17th International Conference on Quality of Multimedia Experience (QoMEX), Madrid, Spain,2025, pp. 1–7. doi: 10.1109/QoMEX65720.2025.11219980. https://ieeexplore.ieee.org/document/10178496}
  
\newcommand\copyrightnotice{%
\begin{tikzpicture}[remember picture,overlay,shift={(current page.south)}]
  \node[anchor=south,yshift=10pt] at (0,0) {\fbox{\parbox{\dimexpr\textwidth-\fboxsep-\fboxrule\relax}{\copyrighttext}}};
\end{tikzpicture}%
}
\copyrightnotice

\section{Introduction}

Recent advancements in computer graphics and animation have led to the prominence of virtual avatars across various domains, including healthcare, education, and entertainment \cite{Norouzi2018-uu}. Depending on their roles, virtual characters have the potential to enhance user experiences in multiple ways. For example, in healthcare applications, they can improve the level of assistance provided to patients and enhance their sense of social presence during interactions \cite{Lee2004}. Avatars can also act as an initial point of contact before a patient is redirected to a human medical professional, potentially fostering a sense of companionship within mental healthcare applications \cite{Emi}. Despite their advantages, the acceptance and effectiveness of virtual avatars in mental healthcare remain areas that require further exploration \cite{shaikh}. Virtual avatars and conversational agents have been employed to assist users in making healthier lifestyle choices and improving mental health \cite{14,30,50}. By assuming human-like responsibilities such as counseling, avatars can improve satisfaction, efficiency, and acceptability through facial, gestural, and verbal cues \cite{37, wang}. However, the impact of avatar design elements, including clothing, hairstyle, ethnicity, and body shape, remains an open research question \cite{56}. Moreover, aspects such as users' identification with self-like avatars and the design attributes of avatars—such as competence, warmth, and friendliness impact Quality of Experience (QoE) and trust to share personal data, as explored in previous research \cite{Curtis2021}, though these aspects are beyond the scope of this work. In the context of healthcare applications, two open questions that remain empirically unaddressed are the avatar size relative to the users and the social distancing.

Photorealistic virtual avatars offer a wide range of human-like facial expressions, body movements, and lip sync, making them suitable for various use cases including healthcare applications. Additionally, AI speech synthesis models can now generate low-latency speech audio from text that is nearly indistinguishable in clarity and expressiveness from actual human speech. The development of photorealistic digital humans has been greatly facilitated by tools such as Epic Games’ MetaHuman Creator \cite{epic}, which allows for the efficient generation of high-fidelity virtual avatars. While the creation of realistic digital humans previously required significant technical expertise and resources, MetaHuman Creator provides a streamlined, accessible solution that enables consistent and detailed avatar design. This technological advancement has made it feasible to incorporate lifelike avatars into research settings, such as our study on user perception in virtual health assistance systems. Despite their growing use, one underexplored aspect is how different sizes of these avatars influence user perception and interaction, particularly in physically immersive environments where avatars are projected onto large surfaces. Understanding how avatar scale affects user experience is essential for optimizing virtual interactions in fields such as virtual healthcare, training simulations, and virtual reality applications. 

In this study, we investigated how the perceived size of MetaHuman avatars affects user interaction and engagement by projecting avatars of varying sizes onto a wall using a projector. Participants interacted with different-sized avatars and provided subjective ratings on various experiential factors, such as realism, presence, and comfort. This approach allowed us to examine how the avatar scale influences user perception and whether life-sized or smaller avatars elicit different psychological and behavioral responses. By analyzing participants’ feedback, we aim to provide insights into the optimal avatar dimensions for enhancing user experience in virtual environments. 

\section{Related Work} \label{literature}


A significant factor influencing the success of virtual avatars in healthcare applications is self-disclosure, which refers to the willingness of individuals to share personal information, thoughts, and experiences to facilitate accurate diagnoses \cite{archer,fiske}. Research indicates that avatars can encourage self-disclosure by leveraging verbal and nonverbal communication cues \cite{33,32}. For example, Moon et al. \cite{moon} demonstrated that self-disclosure by an interviewer prompted participants to share more intimate details. However, open questions remain regarding how avatars can facilitate deep self-disclosure over time \cite{lee}. Warmth and competence are two critical social factors that shape individuals' perceptions of virtual avatars, influencing their trustworthiness and overall acceptance \cite{fiske1}. Although avatars are artificial entities, users often perceive and evaluate them as real social agents \cite{moon,nass}. Avatars that exhibit warmth and competence are generally favored, as they instill a greater sense of credibility and trust \cite{10}. Research has shown that avatar characteristics such as gender, age, attire, and appearance significantly impact trust and user engagement \cite{ter2020you}. Furthermore, Rheu et al. \cite{rheu2021systematic} identified five key design themes that influence trust in conversational agents: social intelligence, voice characteristics, communication style, agent appearance, and non-verbal communication. The effectiveness of virtual avatars in healthcare applications extends beyond trustworthiness to their ability to enhance user experience and engagement. 

Avatar size is a significant design factor in immersive environments, impacting user behavior, perception, and interaction dynamics. The **Proteus effect** reveals that users with taller avatars negotiate more assertively in both virtual and real-world settings, highlighting the lasting influence of avatar size on behavior. Larger avatars are perceived as more dominant and persuasive, altering communication dynamics in augmented reality contexts \cite{Yee2009The, Ratan2019Avatar}.

Aligning an avatar’s size with the user's body enhances the body ownership illusion and sense of presence, resulting in more positive emotional responses \cite{Kim2020Impact}. Larger avatars can also make surrounding objects appear smaller, a phenomenon noted in both adults and children \cite{Keenaghan2022Alice}.

In terms of social dynamics, larger avatars may prompt users to increase interpersonal distance, while equal-sized avatars receive more attention and are viewed as more influential in communication \cite{Ratan2019Avatar}. 

In healthcare, life-sized avatars can improve perceived social presence during online consultations, and dynamic avatars—those that move or gesture—are preferred over static ones, enhancing user experience \cite{Brown2020The}. Slight modifications, like a small increase in limb length, can boost task performance without causing discomfort \cite{Palmer2015How}.
 Moreover, engaging with virtual avatars for patient data collection offers advantages over traditional methods like online forms or paper-based questionnaires \cite{37}. This approach enhances content comprehensibility, particularly benefiting individuals with low cognitive or health literacy, and facilitates more intuitive and open-ended responses \cite{35}. However, despite ongoing research, comprehensive guidelines for designing virtual avatars in eHealth applications remain scarce, with most studies focusing on isolated design elements such as cultural background or agent embodiment \cite{vugt}.

A range of methodologies can be utilized to evaluate the quality of systems from the user's perspective \cite{arndt2017exploring}. Furthermore, the user experience of games, which frequently incorporate numerous avatars, can be systematically measured using subjective evaluation methods \cite{moller2015towards}.
Beyond conventional subjective measurement techniques, the assessment of brain signals has been shown to provide valuable insights into user satisfaction \cite{antons2015neural}. 

While virtual avatars and virtual agents share similarities, they differ in their interaction paradigms. Virtual agents, also referred to as conversational agents, facilitate system-user interactions through text or speech \cite{shaked2017}. These agents may operate based on predefined scripts or artificial intelligence-driven interactions, as seen in voice assistants like Siri and Cortana \cite{sciuto2018}. Virtual agents can be either non-embodied or embodied conversational agents, the latter incorporating non-verbal elements such as facial expressions and gestures to create a more natural interaction experience \cite{cassell2001}. In contrast, virtual avatars act as graphical representations of users within digital environments, allowing interaction through controls such as gamepads \cite{gazzard2009}. They are commonly employed in video games, social media, and immersive virtual applications, where they serve as user-controlled entities rather than autonomous agents. The primary distinction lies in the real-time control of avatars by users, as opposed to interaction with a virtual agent, highlighting the nuanced differences in user engagement within digital spaces.


\section{Methodology}
\begin{figure}[ht]
\centerline{\includegraphics[width=\columnwidth]{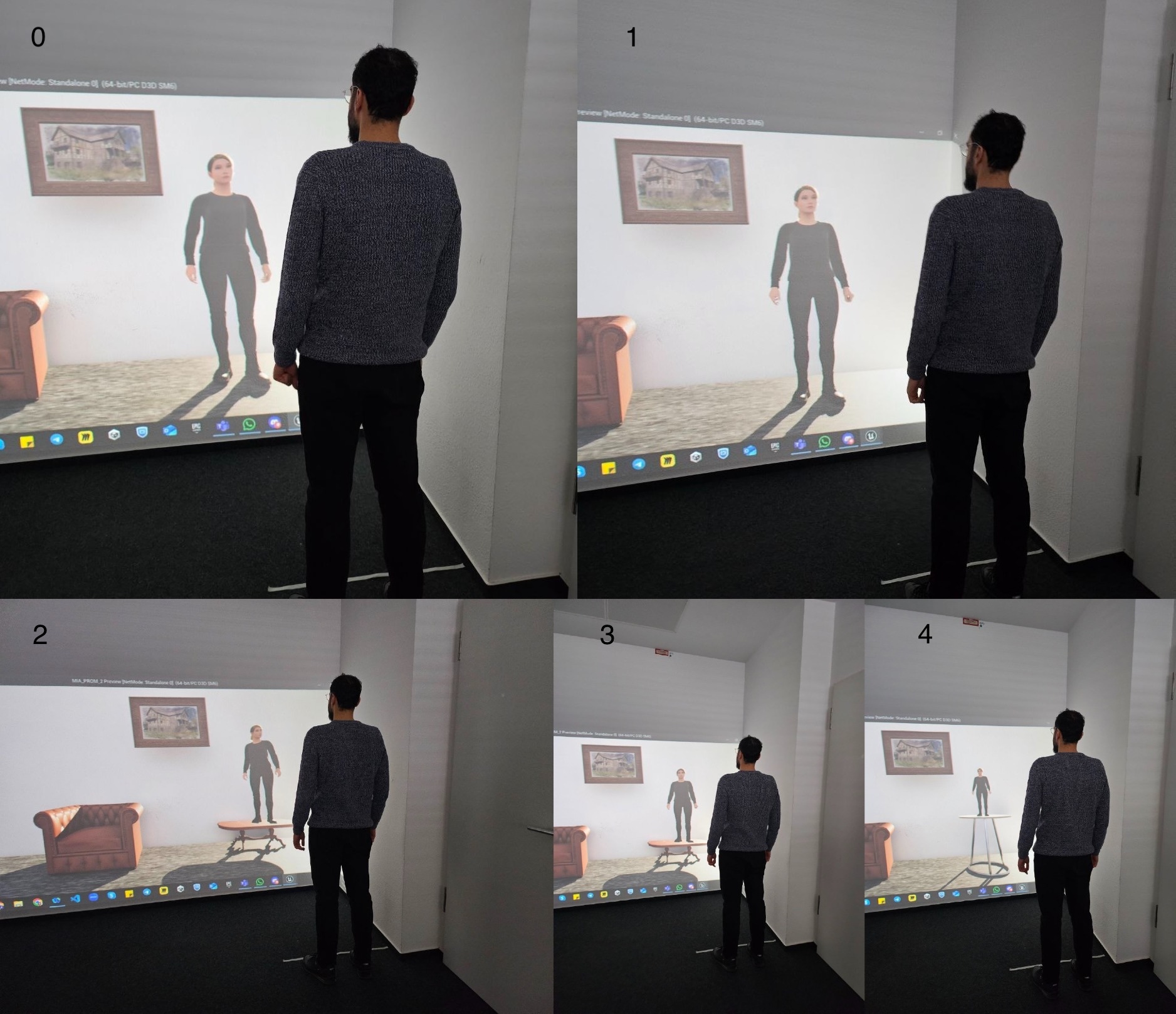}}
\centering
\caption{Experimental setup where users interacted with 5 different avatars from 0 to 4 respectively decreasing in size. (\textit{Image 0 = Human-size avatar: Scale magnitude in Unreal engine: 1 (roughly 175cm in height when projected on the wall), image 1 = Scale magnitude: 0.85 (~150cm), image 2 = Scale magnitude: 0.7 (~130cm), image 3 = Scale magnitude: 0.55 (~100cm), image 4 = Smallest avatar: Scale magnitude: 0.4 (~70cm}).}
\label{avatars}
\end{figure}
\subsection{Experiment setup and proxemic relations}
We developed five Metahuman virtual avatars varying in size for this study, using the Unreal Engine software. Based on the current guidelines and suggestions derived from the literature \cite{appear}, we designed a female avatar with a friendly and competent appearance, a warm and competent voice, and friendly and welcoming facial expressions. We then developed 5 different avatars starting from a human-size avatar (approximately 170cm in height to replicate a standard human size \cite{size}) and four identical avatars with smaller sizes, the smallest of which had a miniature size. The avatars were projected on a wall in landscape mode using a projector with proper placement to avoid users blocking their own view. One challenging task in this study was to position the users in a proper distance from the wall on which the avatars were projected. We have considered the proxemic rules which concern the spatial relations between entities. An extensive study \cite{Hall1966} shows that proxemic relations are mostly implicit and crucial for social order. These culturally specific, institutionalized, and normatively charged rules dictate how individuals position themselves in space. Violations of these rules can provoke strong emotional reactions, making proxemic rules a sensitive yet effective design tool. This has already been recognized in the field of social robotics \cite{Mumm2011, Lauckner2016} and also in the context of virtual-reality \cite{Li-HRIandVRproxemics} or augmented-reality \cite{Huang-ARproxemics} applications, whereas there is currently little work on this in the field of conversational agents. The proposed model by Hall et al. \cite{Hall1966} includes four different proxemic distances: the \textit{intimate space} (within 0.45 meters), \textit{Personal Space} (0.45 to 1.2 meters), \textit{social space} (1.2 to 3.6 meters), and \textit{public space} (above 3.6 meters). We drew a white line approximately 1.5 meters far from the wall for the users to stand behind while interacting with the avatars (Figure \ref{avatars}) to provide a social space distancing according to Hall et al.'s proposal. The examiner have informed the users that the line is only a suggestion and they can stand, sit, or move anywhere surrounding the avatar if they will.  

\subsection{Metahumans and scene design}

The Metahumans were designed using the \textit{Metahuman Creator} tool and imported in the \textit{Ureal Engine} \cite{epic} software for further development. All of the Metahumans were fully integrated with the Convai chatbot \cite{conv} which is a cloud-based AI service provider to make avatars across different platforms conversational by providing a wide range of avatar voices, and facial and body animations. Each avatar was programmed to greet the user and ask five health-related questions chosen from the quality of life questionnaire of the WHO (WHOQOL-BREF) \cite{whool}. To build our scene and background, we used a living room asset environment with vintage furniture in the scene, including couches, tables, and paintings to immersive the user in a calming space. We used the same scene for all avatars, keeping the sofa and painting in the same size to use as references for the users to realize the avatar size difference. Brown et al. \cite{mini} have used miniature-size avatars in a consultant application where avatars would hover in the air to be at the same height has the users' face which has led to a poor QoE. To avoid this scenario, we placed the smaller avatars on furniture items such as tables (Figure \ref{avatars}) to have the avatar face always at the same hight as the user's face. The users were able to communicate to the avatars using a small microphone attached to their shirt. The examiner would activate the recording in a wizard-of-oz manner whenever the user was supposed to speak. We utilized the projector's loudspeaker to propagate the avatar's voice in the room.

\subsection{Experiment procedure and questionnaires}
The users were given an informed consent form to read and sign confirming their voluntary participation in the study, explaining the data privacy policy, the freedom to skip items in the questionnaires, freedom to not answer questions asked by the avatar, and to quit the study at any time if needed. They were then given a verbal description of the study and filled in an Affinity for Technology  Interaction survey (ATI) \cite{ati}, as well as a demographics data questionnaire. The users were then guided by the experimenter to interact with the first avatar. The order of interaction with the avatars was counter balanced. The users would interact with all avatar sizes but in a randomized order. We used a friendly and warm tone of voice offered by Convai for all the avatars to potentially help users bond with and trust the avatars easier. After interacting with each avatar the users were asked to fill in another survey containing a User Experience Questionnaire (UEQ) \cite{ueq} questionnaire (7-point likert scales), UEQ+ on trust also with 7-point likert scales, two additional customized, trust-related questions, and a social presence questionnaire (10-point likert scales) \cite{leee}. We then  asked the participants about their avatar preferences and the reasons behind their choice. The examiner remained in the room for the duration of the experiment to facilitate speech input when needed but did not monitor responses or record any user input during interactions and survey completion.


The implementation of the experiment was approved by the ethics committee Lippstadt\footnote{Ethics processing number: EL202502191}. We used the university recruitment tool to sign participants up for our study. In total, 23 participants took part in the experiment, consisting of 12 males, 10 females, and one not specified (average age 34.69, range 24-65). The user's participation in the study was compensated with 10 Euros in cash. 

\section{Results}

\subsection{Statistical Analysis of Avatar Size Effects}

\begin{figure*}[!h]
    \resizebox{2\columnwidth}{!}{
    \includegraphics[width=0.5\linewidth]{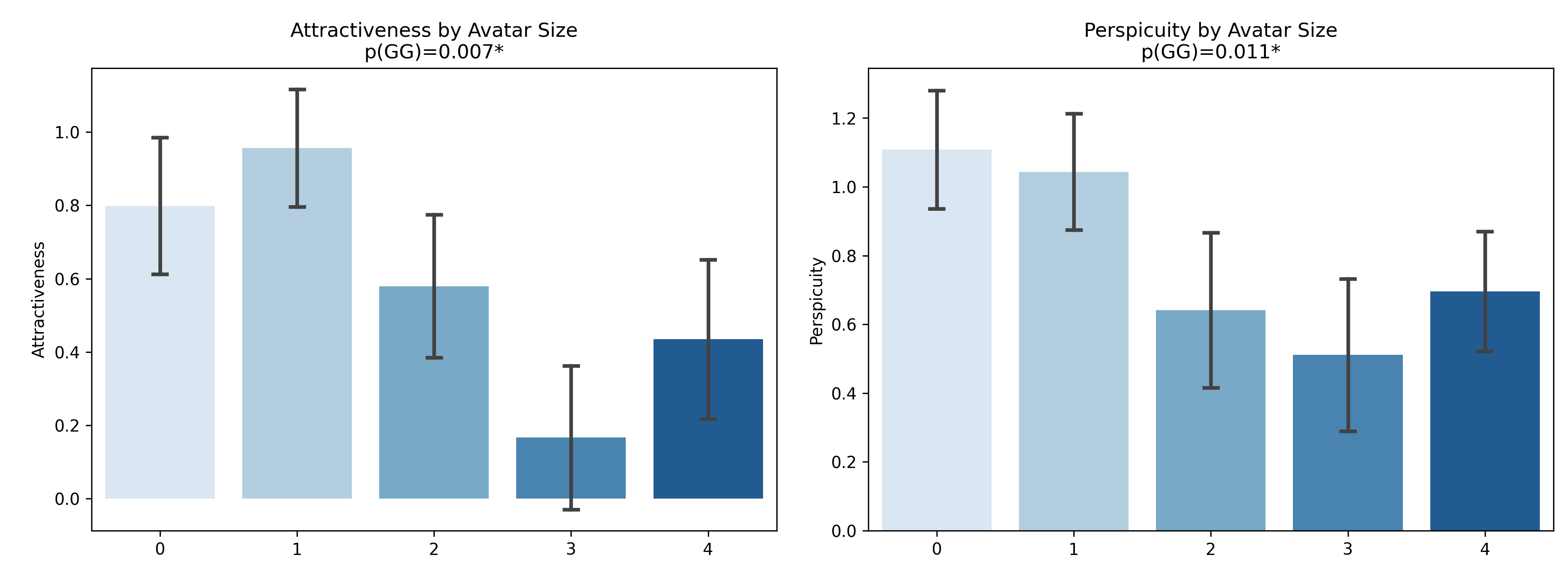}
    }
    \caption{Average UEQ ratings for the avatar sizes (\textit{0 = Human-size avatar, 4 = Smallest avatar}) on the dimensions of \textit{Attractiveness} and \textit{Perspicuity}, with error bars representing the standard error of the mean (SEM).}
    \label{fig:sig_tests}
\end{figure*}

The scale scores for the \textit{UEQ+} were calculated for each participant using the UEQ analysis tool \cite{ueq}, covering the dimensions of \textit{Attractiveness}, \textit{Perspicuity}, \textit{Efficiency}, \textit{Dependability}, \textit{Stimulation}, \textit{Novelty}, and \textit{Trust}. Additionally, an overall score was computed as the average of the seven scales.
An additional trust score was calculated as the mean of two self-formulated items: “\textit{I would be willing to share personal sensitive data with the avatar}” and “\textit{The agent would keep my personal sensitive data secure}”.
Furthermore, a total score was computed for the items of the \textit{Social Presence Questionnaire}.
Subsequently, responses to the five different avatar sizes were compared for the seven $UEQ+$ scales, the overall $UEQ+$ score, the additional trust score, and the \textit{Social Presence Questionnaire} score. Before conducting the analyses, the assumptions for variance analysis were tested using the Shapiro-Wilk test (for normality of residuals) and Levene’s test (for homogeneity of variances). As the normality assumption was violated for \textit{Novelty} and Trust from the $UEQ+$, the additional trust score and the overall $UEQ+$ score, the Friedman test was used instead as a non-parametric alternative. Figure \ref{fig:sig_tests} presents our statistical results.

Among the comparisons, the following results were significant. The Analysis of Variance (ANOVA) for the \textit{Attractiveness} scale revealed a significant difference between avatar sizes, $F(2.86, 62.88) = 4.49$, $p = .007$, $\eta_p^2 = .086$, with a Greenhouse-Geisser correction applied. A post hoc $t$-test with Bonferroni correction identified a significant difference between avatar size 1 and avatar size 3, $t(42.30) = 3.12$, $p =.003$, $p_{adj} = .032$.

An ANOVA also revealed a significant effect of avatar size on \textit{Perspicuity, F(3.46, 76.11) = 3.71}, \textit{p = .011}, $\eta_{\mathit{p}}^2 = .062$, with a Greenhouse-Geisser correction applied. However, the subsequent $t$-test showed a significant difference only between avatar size 0 and avatar size 3, which was no longer significant after Bonferroni correction, $t(41.47) = 2.13$, $p = .039$, $\mathit{p}_{\textnormal{adj}} = .389$.

Additionally, a Fisher’s Exact Test was conducted to examine the association between gender and preferences. The results indicated a significant association, $p < .001$. To assess the effect size, Cram\'{e}r’s V was calculated, yielding a value of $V = 0.52$, indicating a moderate to strong effect.

\subsection{"Corelational, User Perception, and Trust Analysis"}
\textit{Correlation Analysis:} Based on the correlation analysis of our data, several patterns emerge that extend beyond the primary focus on avatar size effects. While avatar size shows consistently weak correlations with all experience metrics (ranging from -0.23 to -0.02), we observed strong inter-correlations between key user experience measures. Social Presence demonstrates strong positive correlations with Stimulation (r = 0.80), Attractiveness (r = 0.75), and Efficiency (r = 0.66). Demographic factors (Age, Gender) and technology affinity showed minimal correlations with experience metrics.


\textit{Effects of ATI Score on Social Presence}: We assessed the relationship between ATI scores and Social Presence across all avatar sizes in the study. The overall correlation between ATI score and Social Presence is positive but weak (r = 0.150) and not statistically significant (p = 0.109). The regression line shows a positive slope, indicating that as ATI scores increase, there is a modest increase in reported Social Presence. The 95\% confidence interval is relatively wide, particularly at the lower end of the ATI scale, which reflects the uncertainty in this relationship and the variability in participant responses.


\textit{Gender-Specific Responses to Avatar Size:} The radar charts reveal distinct gender-specific patterns in participant responses to different avatar sizes. These visualizations were created by calculating mean values for each metric across the five avatar sizes for male and female participants separately, then normalizing these values to a 0-1 scale to facilitate direct comparison while controlling for baseline gender differences.

\begin{figure}[!]
    \resizebox{\columnwidth}{!}{
    \includegraphics[width=0.5\linewidth]{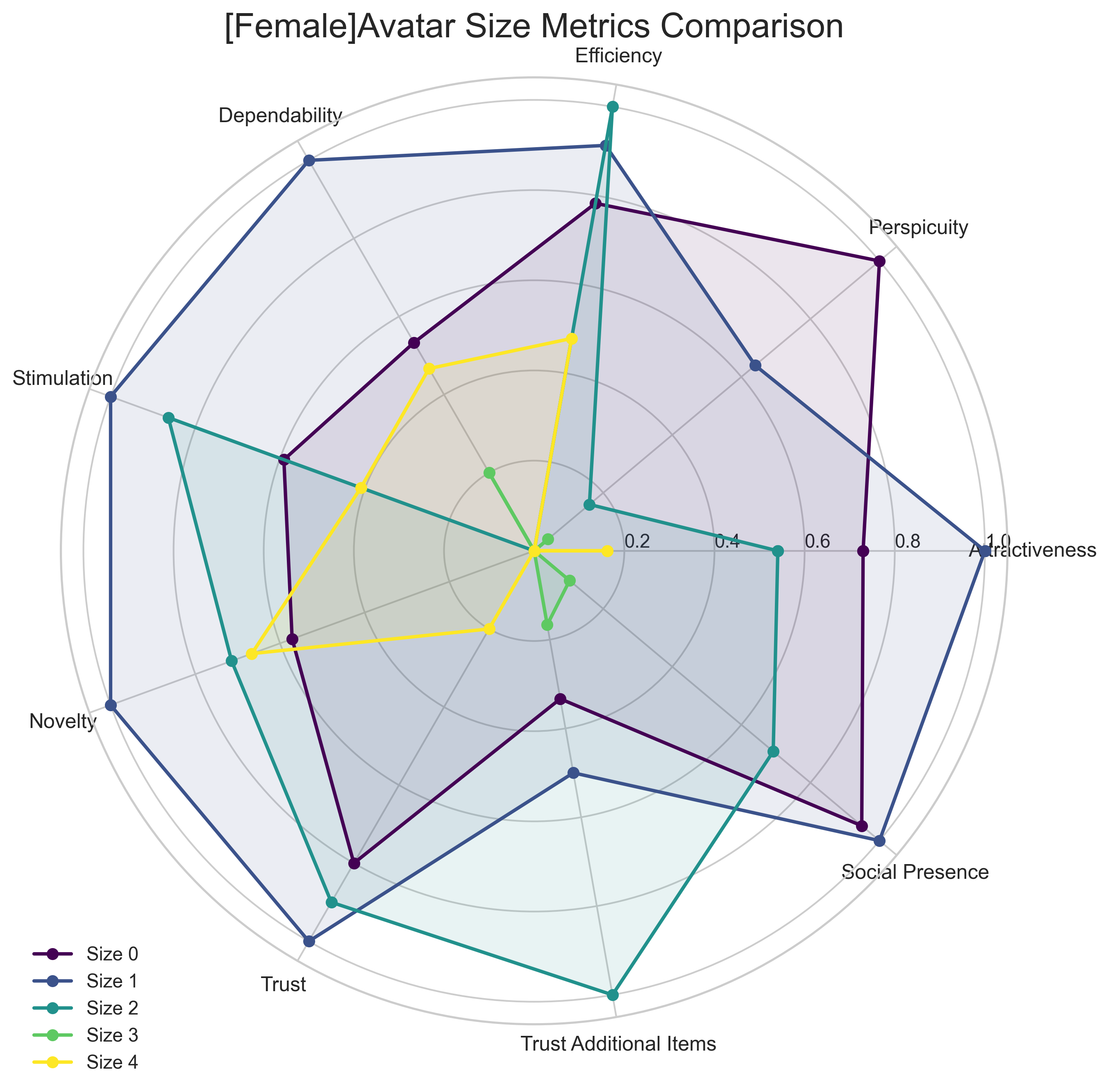}
    }
    \caption{Radar chart depicting average UEQ+ ratings of the female participants for the different avatar sizes (\textit{0 = Human-size avatar, 4 = Smallest avatar}).}
    \label{male}
\end{figure}

\begin{figure}[!]
    \resizebox{\columnwidth}{!}{
    \includegraphics[width=0.5\linewidth]{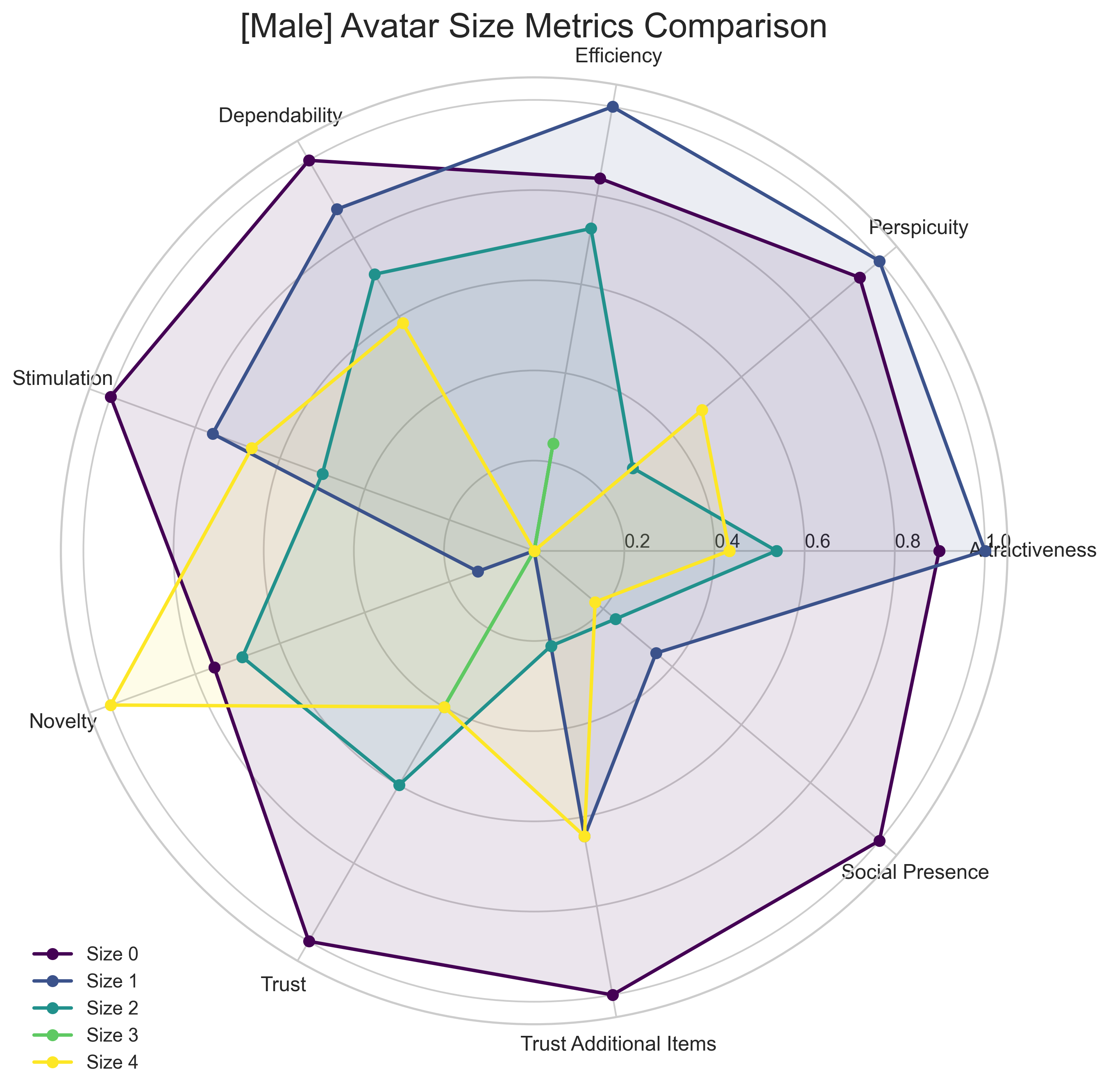}
    }
    \caption{Radar chart depicting average UEQ+ ratings of the male participants for the different avatar sizes (\textit{0 = Human-size avatar, 4 = Smallest avatar}).}
    \label{female}
\end{figure}

Our analysis reveals that female participants responded most positively to Size 1 avatars (slightly smaller than human-sized), particularly regarding Attractiveness, Dependability, and Stimulation (Figure \ref{female}). In contrast, male participants showed stronger preferences for human-sized avatars (Size 0), which received their highest ratings for Attractiveness, Dependability, and Social Presence (Figure \ref{male}). Our results also suggest that the smallest avatar sizes (3 and 4) produced almost inverse patterns between gender groups. Female participants rated Size 4 avatars relatively favorably on trust dimensions compared to Size 3, while male participants showed the opposite pattern.

\section{Discussion}
The results of the ANOVA revealed a significant effect of avatar size on perceived attractiveness, with size 1 being rated as particularly attractive and size 3 as particularly unattractive. Figure \ref{fig:sig_tests} illustrates the relationship between avatar size and attractiveness, showing a tendency for larger avatars to be perceived as more attractive than smaller ones, possibly because they appear more realistic due to their similarity in size to humans. The fact that the second-largest avatar received a higher attractiveness rating than the human-sized avatar may be coincidental, as the difference between these two sizes was not significant. Alternatively, the second-largest avatar might have appeared slightly less intimidating and cuter to participants, leading to a slightly higher rating. This effect could also explain why the smallest avatar received a higher rating than the second-smallest one.

Another ANOVA  test showed a significant effect of avatar size on perceived perspicuity, with a particularly large and initially significant difference between the human-sized avatar and size 3, which, however, was no longer significant after Bonferroni correction. The graphical representation (Figure \ref{fig:sig_tests}) suggests a general trend in which larger avatars received higher perspicuity ratings than smaller ones. It is possible that interacting with these avatars was easier for participants because they are accustomed to similar sizes from human interaction, making them more recognizable, including their mouth movements, which might have facilitated understanding. It is difficult to determine why the smallest avatar, despite this general trend, received a relatively high perspicuity rating (though not significantly higher) compared to sizes 2 and 3. This might have been due to chance, but it is also possible that cuteness effects influenced perspicuity ratings or that smaller avatars evoked associations with characters from video games, making them appear more familiar or easier to interact with.

The strong positive correlations with Stimulation in the social Presence results suggest that these aspects may be more influential in creating a sense of presence than the avatar's size. The particularly strong relationship between Stimulation and Social Presence aligns with our mixed-effects model findings, where Stimulation emerged as the sole significant predictor. Furthermore, the substantial correlations between Attractiveness and other metrics (Efficiency: r = 0.83; Stimulation: r = 0.82) suggest that visually appealing avatars may enhance multiple aspects of the user experience simultaneously. For Size 0 (human representation), a similar weak positive correlation (r = 0.171) is observed between ATI scores and Social Presence. This suggests that the relationship between technology affinity and social presence remains consistent regardless of whether avatars or human representations are used. These findings imply that individual differences in technology affinity may play only a minor role in determining social presence experiences in virtual interactions. 
The non-significant relationship indicates that virtual environments may be accessible to users with varying levels of technology affinity, as higher affinity does not strongly predict better social presence outcomes. This has positive implications for the inclusive design of virtual communication tools.

The differences in the gender-specific response to avatar size may reflect gender-based variations in spatial perception and social interaction expectations documented in previous virtual environment research mentioned in \ref{literature}. The non-linear relationship found between the gender groups suggests that avatar size optimization should consider target audience characteristics rather than assuming uniform preferences. These findings extend our understanding beyond the main ANOVA results by highlighting how demographic factors can moderate the relationship between avatar size and user experience. The observed gender-specific patterns emphasize the importance of considering individual differences when designing virtual characters, as optimal avatar size may vary depending on user characteristics. Additionally, the observed patterns may be relatively consistent across different user characteristics when considering the demographics and ATI scores. These findings collectively suggest that the quality of interaction and the avatar's appeal may outweigh its size in determining how users experience virtual social presence. While there might be a slight tendency for participants with higher technology affinity to experience greater social presence in virtual environments, this relationship is not strong enough to be considered reliable.

\section{Conclusion}
In this work, we assessed user perception and preferences when interacting with highly realistic Metahuman avatars projected on a wall at different sizes. These avatars were fully integrated with an AI chatbot to enable human-like conversations, supported by lifelike body and facial animations. Our findings suggest that avatar size significantly influences user perception—particularly in terms of attractiveness and perspicuity—with human-sized and larger avatars generally rated higher than smaller ones across usability, trust, and social presence dimensions. Gender differences emerged, with women tending to prefer human-sized avatars and men exhibiting more varied preferences. Correlation analyses revealed strong associations between social presence, stimulation, and attractiveness, highlighting the key role of visual appeal in shaping user experience. While technology affinity showed only weak effects on social presence, the gender-specific responses underline the importance of tailoring avatar design to different user groups. Although our sample size (n = 23) offers valuable initial insights, it is not sufficient to draw definitive conclusions. 

In response to display modality, our decision to project avatars onto a wall—rather than use head-mounted displays—was guided by practical considerations and offers a unique, non-intrusive interaction setup, especially valuable in scenarios such as healthcare or public installations where HMDs may be unsuitable. As for avatar scale, the study focused on human-sized and smaller avatars based on the premise that larger-than-life representations could appear unnatural or overwhelming in face-to-face conversational contexts. However, future investigations could explore exaggerated avatar sizes in specific narrative or entertainment settings to assess whether such designs could enhance engagement or immersion under different conditions. Overall, our findings contribute to the growing body of research on digital human interaction and offer practical implications for avatar design in virtual healthcare settings, particularly with respect to proxemics and perceived realism. Continued exploration of avatar scale, emotional adaptability, and contextual fit will further improve the quality of experience in avatar-based applications.

\section*{Acknowledgment}
The authors used ChatGPT4.0 and Grammarly for grammar and spell-checking. After using these tool(s)/service(s), the authors reviewed and edited the content as needed and take full responsibility for the publication’s content.

\bibliographystyle{IEEEtran}
\bibliography{conference_101719.bib}

\end{document}